%%%%%%%%%%%%%%%%%%     tex file       %%%%%%%%%%%%%%%%%%%%%%%%%%%%%%%%%
%%%%%%%%%%%%%%%%%%%%%%%%%%%%%%%%%%%%%%%%%%%%%%%%%%%%%%%%%%%%%%%%%%%%%%%
\input harvmac
%\draftmode
\Title{hep-th/9707036}
{\vbox{\centerline{'t Hooft Anomaly Matching Conditions}
\bigskip \centerline{for Generalized Symmetries in 2D}}}
\bigskip
\centerline{F. Bastianelli $^{a,b,}$
\footnote{$^*$}{email: bastianelli@imoax1.unimo.it}
\  and \  O. Corradini $^{a,}$
\footnote{$^\dagger$}{email: olindo@unimo.it}}
\centerline{$^{a}$ Dipartimento di Fisica,
Universit\`a di Modena, via Campi 213/A, I-41100, Modena,}
\centerline{and}
\centerline{$^{b}$ INFN, Sezione di Bologna, Bologna, Italy.}
\vskip 2cm

\noindent
The 't Hooft anomaly matching conditions are a standard tool to 
study and test non-perturbative issues in quantum field theory.
We give a new, simple proof of the anomaly matching conditions
in 2D Poincar\`e invariant theories. We consider the case
of invariance under a large class of generalized symmetries,
which include abelian and non-abelian  internal symmetries,
space-time symmetries generated by the stress tensor, and 
$W$-type of symmetries generated by higher spin currents.

\Date{6/97}

\newsec{Introduction}

The importance of the anomaly matching conditions has been 
recognized by 't Hooft in a classic paper 
\ref\thooft{G. 't Hooft, in \lq\lq Recent Developments in Gauge
Theories", p. 135, G. 't Hooft et al. eds. 
(Plenum Press, New York, 1980).} as a non-perturbative tool to 
put restrictions on the quantum numbers of 
composite massless particles.
Since then the anomaly matching conditions have been used
as a major tool to search for preonic models
\ref\mars{see e.g., R. Marshak, 
\lq\lq Conceptual Foundations of Modern 
Particle Physics\rq\rq\ (World Scientific, Singapore, 1993).},
and to analyze strongly coupled N=1 supersymmetric models 
\ref\seiberg{
N.Seiberg, hep-th/9402044, Phys. Rev. D49 (1994) 6857\semi
N.Seiberg, hep-th/9411149, Nucl. Phys. B435 (1995) 129\semi
for a review see: K. Intriligator and N. Seiberg, hep-th/9509066.
Nucl. Phys. Proc. Suppl. 45BC (1996) 1.}.
Their power is related to the fact that they constitute an exact,
fully non-perturbative result obtained in the framework of 
quantum field theory.
As it is familiar to field theory practitioners,
there are not so many exact results known for 
interacting quantum field theories.

The arguments given by 't Hooft for the anomaly matching 
have been analyzed and extended in several papers 
\ref\fsby{Y. Frishman, A Schwimmer, T. Banks and S. Yankielowicz, 
Nucl. Phys. B177 (1981) 157\semi
S. Coleman and B. Grossman, Nucl. Phys. B203 (1982) 205.},
and by now they stand on a rather solid footing. 
Nevertheless, it may be useful
to study the matching conditions from different perspectives.
It is the purpose of this letter to give a new, simple proof of 
such an important result
in the simple context of two dimensional theories.
We show that the coefficients related to the anomalies are 
invariant
under the flow generated by the renormalization group (RG).
We will employ a method of proof similar to that originally 
introduced by
Zamolodchikov in his study of the $C$-theorem
\ref\Z{ A.B. Zamolodchikov, JETP Lett. 43 (1986) 730;
Sov. J. Nucl. Phys. 46 (1987) 1090.}.

An extension of the $C$-theorem to chiral theories was already 
studied in 
\ref\BL{F. Bastianelli and U. Lindstr\"om,
hep-th/9604001, Phys. Lett. B380 (1997) 341.},
where it was noted that a certain function was left invariant 
by the action of the renormalization group, 
and related to the gravitational anomalies
\ref\AGW{L. Alvarez-Gaum\'e and E. Witten, 
Nucl. Phys. B234 (1984) 269.}.
In this paper we use a similar philosophy, 
and search for functions that are constant along the RG group 
trajectories. Then, we observe that these functions are related 
to chiral anomalies.
The values of these chiral anomalies are easily identified
as a combination of the left and right
central charges of the corresponding  conformal symmetry algebra.

The physical interpretation is quite clear. The anomaly 
coefficients are seen to be invariant along the trajectories  
generated by the beta functions of the theory:
the values of the coefficients
generated by the microscopic degrees of freedom, visible in the 
ultraviolet region,  must be reproduced by the macroscopic degrees 
of freedom which describe the infrared physics.

We will proceed as follows.
In sec. 2 we prove the 't Hooft anomaly matching conditions 
for the case of internal abelian and non-abelian symmetries.
In sec. 3 we extend the proof to the case of
generalized symmetries generated by spin-$n$ currents. 
For $n=2$ this reproduce the case of the stress tensor.
In sec. 4 we consider mixed anomalies.
Finally, in sec. 5 we present our conclusions.

\newsec{The case of global internal symmetries}

Let us consider the case of a two dimensional quantum field theory 
invariant under an abelian $U(1)$ global symmetry. 
The internal symmetry is generated by a conserved current $J_\mu$: 
$\partial^\mu J_\mu=0$.
Using complex coordinates\footnote{$^1$}{We work in the euclidean 
version of the theory by performing a
Wick rotation. Nevertheless, we use a language 
appropriate to the minkowskian theory.},
we denote the independent components of the symmetry current
by $J = J_{z}$, $\bar J = J_{\bar z}$, so that 
the conservation equation reads
\eqn\uno{\bar \partial J + \partial \bar J =0.}
In a Poincar\`e invariant quantum field theory, the two-point 
functions 
of the symmetry current must have the following general form
\eqn\due{\eqalign{ 
<J(z,\bar z) J(0,0)> &=  {K(z \bar z)\over z^2},\cr
<J(z,\bar z) \bar J(0,0)> &= {H(z \bar z)\over {z \bar z}}, \cr
<\bar J(z,\bar z) \bar J(0,0)> &= 
{L(z \bar z)\over {\bar z^2}},\cr}}
where $K,H,L$ are undetermined scalar function of the 
product ${z \bar z}$.
In fact, translation invariance implies that the two-point
correlation functions depend only on the relative distance 
between the 
points, and one can use this invariance to fix the coordinate of
the second point at the origin
of the coordinate system.
Lorentz covariance is then used to extract the 
expected Lorentz transformation properties. Moreover,  
since conserved currents do not develop anomalous dimensions, 
the scalar functions $K,H$ and $L$ must be dimensionless,
since the expected dimensions are already carried by the 
$z$ and $\bar z$ dependence factored out.
Imposing the conservation equation \uno\ onto \due,
one obtains the following relations 
\eqn\tre{ 
\bar\partial \biggl({K\over {z^2}} \biggr)
 + \partial \biggl({H\over {z \bar z}}\biggr)=0, \ \ \ \
\partial \biggl({L\over {\bar z^2}}\biggr) + 
\bar\partial \biggl({H\over {z\bar z}} \biggr)=0, }
which can be rewritten as
\eqn\quattro{ r^2{\partial  \over \partial r^2}
(K+H) = H , \ \ \ \ 
 r^2{\partial  \over \partial r^2}
(L+H) = H  }
where $ r^2 = z \bar z$.
Note that we consider only non-vanishing finite distances, 
$r \neq 0$,
so that it is consistent to drop possible contact terms from
eqs. \due\ and \tre\
(i.e. terms proportional to delta functions and derivatives 
thereof: indeed
such terms are generically present in non-trivial 
quantum field theories).
From eq. \quattro\
it is immediate to recognize that the quantity
\eqn\quattrob{ A \equiv K-L }
satisfies
\eqn\cinque{ {\partial A \over \partial r^2}  =0.}
Thus, we see that the \lq\lq anomaly coefficient" $A$ is 
independent of the distance scale $r$.
This is the essence of the 't Hooft anomaly matching conditions: 
the anomaly coefficient $A$ is constant over the various length 
scales. To complete the argument,
one can relate this constancy to a constancy along
the trajectory generated by the renormalization group flow.
We use dimensional analysis
\eqn\cinquea{
\biggl (\mu {\partial\over {\partial  \mu}} -
2 r^2 {\partial \over \partial r^2} \biggr) A = 0, } 
and the renormalization group equation
\eqn\cinqueb{
\biggl(\mu {\partial\over {\partial \mu}} +
\beta^i {\partial \over {\partial g^i}} \biggr) A = 0,}
where $\mu$ is the mass scale parametrizing the choice of the
renormalization conditions, and 
$\beta^i = \beta^i (g^j)$ are the beta functions of the theory.
As usual, the beta functions can be integrated to obtain the
running coupling constants $g^i(t)$ 
\eqn\cinquec{ {d\over {d t}}
g^i(t)= - \beta^i (g^j). }
Now, we can compute the variation of $A$ along the trajectory 
generated by the renormalization group flow
\eqn\sei{ {d\over {d t}} A 
\equiv -\beta^i {\partial \over {\partial g^i}} A 
= \mu {\partial\over {\partial \mu}} A
= 2 r^2 {\partial\over {\partial r^2}} A =  0.}
This proves the 't Hooft anomaly matching conditions.
We see that the function $A$ is constant along
the trajectories generated by the renormalization group flow:
its value at the ultraviolet fixed point must be reproduced by 
whatever
macroscopic degrees of freedom describe the low-energy physics.

Now, let us relate the function $A$ to the properties 
of the conformal field theories describing the 
ultraviolet (UV) and infrared (IR) fixed points
which characterize the ending points of the RG trajectory.
At these fixed points conformal invariance takes over,
and the equation \uno\ split into two independent pieces 
generating the left and right moving current algebras
\eqn\sette{\eqalign{
\bar  \partial J = 0 , \ \ \ \ \ \ \ &J(z) J(0) = 
{k \over {z^2}}, \cr
\partial \bar J = 0, \ \ \ \ \ \ \
&\bar J(\bar z)\bar  J(0) = {\bar k \over {\bar z^2}}, \cr }}
where $k$ and $\bar k$ are the central charges
for the left and right moving $U(1)$ currents.
From eq. \quattrob, and taking into account eq. \sei, we can
immediately read off the following equalities
\eqn\otto{ A= k_{_{UV}} - \bar k_{_{UV}} =   
k_{_{IR}} - \bar k_{_{IR}},}
where $ (k_{_{UV}},\bar k_{_{UV}})$ and 
$( k_{_{IR}} , \bar k_{_{IR}})$ denote the left and right
central charges of the UV and IR conformal current algebras,
respectively. This result justify the name  
``anomaly coefficient'' for $A$. 
In fact, the conserved current $J_\mu$ can be 
coupled to a gauge field $A_\mu$ in a gauge invariant 
way only if $A=0$.

The previous analysis is easily extended to the case of an 
internal 
simple symmetry group $G$ generated by conserved currents 
$J^a_\mu$:
$\partial^\mu J^a_\mu=0, \ \ a=1,.., {\rm dim}\  G$.
One can use the additional invariance under the group $G$ to 
constrain the two-point functions of the symmetry currents
\eqn\nove{\eqalign{
<J^a(z,\bar z) J^b(0,0)> &=  
{ \gamma^{ab} K(z \bar z)\over z^2},\cr
<J^a(z,\bar z) \bar J^b(0,0)> &= {\gamma^{ab}
H(z \bar z)\over {z \bar z}}, \cr
<\bar J^a(z,\bar z) \bar J^b(0,0)> &= {\gamma^{ab}
L(z \bar z)\over {\bar z^2}},\cr}}
where $\gamma^{ab}$ is the Killing metric of the group $G$.
One can check that the same function defined in \quattrob\
is invariant along the RG trajectory.
It is related to the difference
of the central charges (the \lq\lq levels") appearing 
in the Kac-Moody 
algebras which characterize the UV and IR fixed points.
For completeness, we recall that a Kac-Moody
algebra is described by the following
operator product expansion
\eqn\km{ J^a(z)  J^b(0) = {k \gamma^{ab} \over z^2} + 
{i f^{ab}{}_{c} J^c(0) \over z}}
where the central charge $k$ is called the level, and where 
$\gamma^{ab}$ and $f^{ab}{}_{c}$ denote the Killing metric and 
the structure 
constants of a simple Lie group $G$, respectively.

\newsec{The case of generalized symmetries}

We consider now a theory which is invariant under symmetries 
generated by a spin-$n$ conserved current 
\eqn\duno{\partial^{\mu_1} W_{\mu_1 \mu_2 ... \mu_n} = 0 ,}
where $W_{\mu_1 \mu_2 ... \mu_n}$ is 
a completely symmetric tensor.
A class of 2D theories invariant under such 
generalized space-time symmetries have been first
identified by Zamolodchikov 
\ref\zam{ A.B. Zamolodchikov, Theor. Math. Phys. 63 (1985) 1205\semi
V.A. Fateev and S.L. Lukyanov, 
Int. J. Mod. Phys. A3 (1988) 507\semi
for a review see: 
P. Bowkgnet and A. Schoutens, 
hep-th/9210010, Phys. Rep. 223 (1993) 183.}
as particular examples of conformal field theories 
with higher spin symmetry currents, 
generating the so-called $W$-algebras.
In general, the current in eq. \duno\ will generate 
off-critical space-time symmetries.

We use again complex coordinates,
and denote by $W_p$ the component of \duno\ with 
$p$ holomorphic indices,
since then the number of antiholomorphic indices is uniquely
fixed to be $n-p$ (e.g.
for $n=3$ we denote $W_3 \equiv W_{zzz}$, 
$W_2 \equiv W_{zz\bar z}$, $W_1 \equiv W_{z \bar z\bar z}$,
$W_0 \equiv W_{\bar z \bar z \bar z}$).
In this notation the conservation equations read 
\eqn\ddue{\bar \partial W_p + \partial W_{p-1} =0, 
\ \ \ \ \ \ \ \ \ p=1,..,n.}
By using Poincar\`e covariance, the two-point functions of the 
current are seen to have the following general  form
\eqn\dtre{
<W_p(z,\bar z) W_q(0,0)> =  {{F_{p,q}(z \bar z)}\over {z^{p+q} 
\bar z^{2n-p-q}}}
, \ \ \ \ \ \ 0\leq p,q \leq n.}
The scalar functions $F_{p,q}$ form a symmetric matrix
with ${(n+1)(n+2)\over 2}$ components, but not all of its 
components are independent, as we shall see later on.
At a critical point only the $F_{n,n}$ and $ F_{0,0}$ 
components are non-vanishing, and they will be 
related to the central charges of the corresponding $W$-like 
conformal algebras.

Imposing the conservation equation on the two-point functions,
one deduces the following equations
\eqn\dtre{ r^2{\partial  \over \partial r^2}
(F_{p,q} + F_{p-1,q}) = (2n-p-q)F_{p,q} + (p+q-1) F_{p-1,q} ,}
where we continue to denote $r^2 = z \bar z .$
To search for a constant function, one can start from
\eqn\dtreb{
r^2{\partial  \over \partial r^2} (F_{n,n} + F_{n-1,n})
= (2n-1) F_{n-1,n} }
and subtract from it a similar equation to eliminate 
the $F_{n-1,n}$ dependence on the right hand side.
This is achieved by subtracting a term proportional to
\eqn\dtret{
r^2{\partial  \over \partial r^2} (F_{n-1,n} + F_{n-1,n-1})
= F_{n-1,n} + (2n-2) F_{n-1,n-1} .}
Then, in a stepwise fashion one tries to reach $F_{0,0}$
which satisfies
\eqn\dtreq{r^2{\partial  \over \partial r^2} 
(F_{0,0} + F_{1,0})
= (2n-1) F_{1,0} .}
Proceeding this way, one is led to consider the function
\eqn\dquattro{\eqalign{ A =
&\sum_{k=1}^{n} \biggl [ {\pmatrix{ 2n-1 \cr 2k-2}} 
(F_{n-k+1,n-k+1}  + F_{n-k,n-k+1} ) 
\cr &- 
\pmatrix{2n-1 \cr 2k-1}
(F_{n-k,n-k} + F_{n-k+1,n-k} ) \biggr ],}}
which indeed satisfies 
${\partial\over {\partial r^2}} A =0$.
The anomaly $A$ can also be written in the following 
somewhat more compact form
\eqn\dsei{\eqalign{ A = F_{n,n} - F_{0,0} &+
\sum_{k=1}^{n-1} {n-2k\over {n-k}} \pmatrix{2n-1 \cr 2k}
F_{n-k,n-k} \cr & -
\sum_{k=1}^{n} {2n-4k+2 \over {2n-2k+1}} 
\pmatrix{2n-1 \cr 2k-1}
F_{n-k,n-k+1} .\cr}}
At a critical point the only non-vanishing functions will be 
$F_{n,n}$ and $F_{0,0}$, so that the anomaly 
reduces to $A=F_{n,n} - F_{0,0}$,
where $F_{n,n}$ and $F_{0,0}$ are related to the left and right 
central charges appearing in the critical $W$-like algebras.
For $n=2$, this reproduces the case of the stress tensor
$T_{\mu\nu}$, and the anomaly matching 
$2 A= c_{_{UV}} -\bar c_{_{UV}} = c_{_{IR}} -\bar c_{_{IR}}$
gives the matching of the gravitational anomalies 
($c$ and $ \bar c$ denote the left and right
Virasoro central charges, respectively).

We must note that for the special case of higher spin
symmetry currents, when the fixed point
algebras reduces precisely to those discovered by 
Zamolodchikov and Fateev-Lukyanov, the anomaly matching 
does not give a new independent condition. In fact, the 
central charges $F_{n,n}$ and $F_{0,0}$ are linearly related 
by the Jacobi identities to the left and right
central charges appearing in the sub-algebra generated by
the stress tensor, the Virasoro algebra.
The latter is always present since we consider Poincar\`e
invariant theories.

In more general cases,  when the current
$W_{\mu_1 \mu_2 ... \mu_n}$ reduces at the fixed points to
currents generating conformal 
algebras with independent central charges, or 
at least with non-linear relations
between the various central charges, the constancy of $A$
gives new independent anomaly matching conditions.

Finally, we note that there are many equivalent ways 
of writing the anomaly coefficient $A$ in \dsei. By repeated use
of the conservation equation \ddue, one derives the identities
$F_{p,q-1} = F_{p-1,q}$, valid for $p+q \neq n+1$.
These identities may be used to 
cast the anomaly $A$ in different looking forms.

\newsec{Anomaly matching for mixed anomalies}

Whenever there is a central charge appearing in the conformal 
algebra describing a fixed point of the renormalization group, 
one can derive a matching conditions for those massive theories 
connected to this critical point by a RG group trajectory.
What one needs to prove is the invariance along the 
RG trajectory, 
starting or ending at the given fixed point,  
of a combination of the left and right central charges.
This may be called \lq\lq anomaly matching for mixed anomalies".

To be concrete, let us consider a particular example of a 
conformal algebra describing a fixed point theory. 
We take for the left moving sector 
\eqn\quno{\eqalign{
T(z) T(0) &= {c \over {z^4}} + {2 T(0) \over z^2}
+{ \partial T(0) \over z} , \cr
T(z)J(0) &= { c' \over {z^3}}+ {J(0) \over z^2}+
{ \partial J(0) \over z} , \cr 
J(z) J(0) &= { k \over {z^2}}, \cr }}
and a copy of a similar algebra characterized by 
the central charges
$(\bar c, \bar c', \bar k)$ for the right moving sector.
This algebra may be considered as the conformal algebra 
describing the (UV or IR) fixed point of certain massive 
2D theories 
with a conserved stress tensor and a conserved $U(1)$ current.
We have already seen how to derive an anomaly matching 
for $c-\bar c$ and $k-\bar k$. Using the same type of procedure
described in the previous sections, 
it is quite easy to derive a matching 
condition for the mixed anomaly $c' + \bar c'$.
We introduce the general two-point functions
\eqn\qdue{\eqalign{
<J(z,\bar z) T(0,0)> &=  {F_1(z \bar z)\over z^3},\ \ \ 
<\bar J(z,\bar z) T(0,0)> = {F_2(z \bar z)\over {z^2 \bar z}}, 
\cr 
<\bar J(z,\bar z) \Theta (0,0)> &= 
{F_3(z \bar z)\over {z \bar z^2}},\ \ \
<J(z,\bar z) \Theta(0,0)> =  
{F_4(z \bar z)\over z^2 \bar z},\cr
<J(z,\bar z) \bar T(0,0)> &= 
{F_5(z \bar z)\over {z \bar z^2}}, \ \ \
<\bar J(z,\bar z) \bar T (0,0)> = 
{F_6(z \bar z)\over {\bar z^3}},\cr
}}
where $\Theta \equiv T_{ z\bar z}$.
Imposing the conservation equations
\eqn\ttre{\bar \partial J + \partial \bar J =0, \ \ \ \ 
\bar \partial T + \partial \Theta =0,\ \ \ \
\partial \bar T + \bar \partial \Theta =0,}
one can derive 
\eqn\tqua{{d\over dr^2}A = 0, \ \ \ \ \ \ A 
\equiv F_1-F_2 -F_5 +F_6.}
(Note that one can also prove the relations
$F_2=F_4 $ and $ F_3 = F_5$).
Then, by using the RG equations, one obtains the required anomaly 
matching conditions
\eqn\tcinque{ {d\over {d t}}A
\equiv -\beta^i {\partial \over {\partial g^i}} A =  0 \ \ \ \ 
\Longrightarrow \ \ \ \
A = c'_{_{UV}} + \bar c'_{_{UV}} =
c'_{_{IR}} + \bar c'_{_{IR}}.}

The general case is treated in a similar way: 
if a central charge $c$ appears in the operator product 
expansion of a spin-$m$ and a spin-$n$
symmetry currents, then one can derive an anomaly 
matching condition for $A= c - (-1)^{m+n}\bar c$.

\newsec{Conclusions}

We have described a simple method for proving the 't Hooft
anomaly matching conditions in 2D quantum field theories, and 
considered theories which may be invariant under a large class of 
generalized symmetries.
We have employed a method of proof similar to that used by 
Zamolodchikov for obtaining the $C$-theorem.
We may note that in our case unitarity was not required
as it was in the proof of the $C$-theorem. Therefore,
we see that the anomaly matching is still valid
in non-unitary Poincar\`e invariant theories.
This fact may be of relevance for applications in condensed 
matter systems (e.g. as in 
\ref\andrei{N. Andrei, M.R. Douglas A. Jerez, cond-mat/9502082
``Chiral Non-Fermi Liquids in 1-d''.}).

Both the $C$-theorem and the 't Hooft anomaly matching conditions
are quite powerful non-perturbative results obtained in 2D
quantum field theories. The extension of the $C$-theorem to four 
dimensions has not been achieved, yet, even though many proposals 
have been analyzed 
\ref\cth{J.L. Cardy, Phys. Lett. B215 (1988) 749\semi
H. Osborn, Phys. Lett. B222 (1989) 97\semi
I. Jack and H. Osborn, Nucl. Phys. B343 (1990) 647\semi 
A. Cappelli, D. Friedan and J.I. Latorre, Nucl. Phys. B352
(1991) 616\semi
G. Shore, Phys. Lett. B253 (1991) 380; B256 (1991) 407\semi
A. Cappelli, J.I. Latorre and X. Vilas\'\i s-Cardona, 
hep-th/9109041, Nucl. Phys. B376 (1992) 510\semi
D. Anselmi, M. Grisaru and A. Johansen, hep-th/9601023, 
Nucl Phys. B491 (1997) 221\semi
D. Anselmi, D.Z. Freedman, M. Grisaru and A. Johansen, 
hep-th/9608125, Phys. Lett. B383 (1996) 415.}. Actually, 
certain tests indicate that such an extension
may be valid, at least in supersymmetric theories 
\ref\fb{F. Bastianelli, hep-th/9511065, Phys. Lett. 
B369 (1996) 249.}.
On the other hand, the 't Hooft anomaly matching conditions
are certainly valid in 4D. Therefore, it may 
be possible to prove them in a way similar to that 
described here.

\listrefs
\end